\documentclass[conference]{IEEEtran}
\IEEEoverridecommandlockouts
\usepackage{cite}
\usepackage{amsmath,amssymb,amsfonts}
\usepackage{algorithmic}
\usepackage{graphicx}
\usepackage{textcomp}
\usepackage{xcolor}
\def\BibTeX{{\rm B\kern-.05em{\sc i\kern-.025em b}\kern-.08em
    T\kern-.1667em\lower.7ex\hbox{E}\kern-.125emX}}
\begin{document}

\newcommand{\linebreakand}{%
  \end{@IEEEauthorhalign}
  \hfill\mbox{}\par
  \mbox{}\hfill\begin{@IEEEauthorhalign}
}

\title{Artificial Neural Networks to Recognize Speakers Division from Continuous Bengali Speech\\
}

\author{
Hasmot Ali\textsuperscript{1}, 
Md. Fahad Hossain\textsuperscript{1}, 
Md. Mehedi Hasan\textsuperscript{1}, 
Sheikh Abujar\textsuperscript{1,2}, 
Sheak Rashed Haider Noori\textsuperscript{1}\\
\textsuperscript{1}Daffodil International University, Dhaka, Bangladesh\\
\textsuperscript{2}The University of Alabama at Birmingham, AL, USA\\
\small \{hasmot15-9632, fahad15-9600, mehedi15-9804\}@diu.edu.bd, sabujar@uab.edu, drnoori@daffodilvarsity.edu.bd
}

\maketitle

\begin{abstract}
Voice-based applications are ruling over the era of automation because speech has a lot of factors that determine a speaker’s information as well as speech.  Modern Automatic Speech Recognition (ASR) is a blessing in the field of Human-Computer Interaction (HCI) for efficient communication among humans and devices using Artificial Intelligence technology. Speech is one of the easiest mediums of communication because it has a lot of identical features for different speakers. Nowadays it is possible to determine speakers and their identity using their speech in terms of speaker’s recognition. In this paper, we presented a method that will provide a speaker's geographical identity in a certain region using continuous Bengali speech. We consider eight different divisions of Bangladesh as the geographical region. We applied the Mel Frequency Cepstral Coefficient (MFCC) and Delta features on an Artificial Neural Network to classify speakers’s division. We performed some preprocessing tasks like noise reduction and 8-10 second segmentation of raw audio before feature extraction. We used our dataset of more than 45 hours of audio data from 633 individual male and female speakers. We recorded the highest accuracy of 85.44\%. 
\end{abstract}


\section{Introduction}
Speech is a powerful medium for efficient communication. Having a bunch of unique features like Qualities, Pitch, Tone, Rhythm, Resonance, Texture, and others, speech is one of the most useful identifiers to recognize speakers. To differentiate speakers from the same language and different places speech features work as a powerful attribute and they work as a unique way of introducing a group of people who speak the same language. People from a geographical area or a community have some unique speech properties like they speak in a specific accent and rhythm. Voice also can be the recognition apparatus of Gender, Occupation, Height, Age, Division, Weight, or any other identical parameter of a speaker.\\

For having perfectly structured multiple-level regularized properties speech can produce the basic unit of speech which is defined as phonemes. It can determine the speech clearly and perfectly but for our contribution, speech features define the nature of speech. By which the identification or characteristic of a specific speaker could be possible easily. MFCC provides the necessary feature to train our model so that the model can determine an unknown speaker’s division. Accents could play an important role in determining people from specific geographical areas. Accent-based speaker recognition is one of the emerging topics for ASR researchers. Deep learning algorithms like RNN, ANN, CNN, and LSTM are performing better for speech recognition because of their perfectly structured data. For our data, ANN performs best with the system and configuration we used.\\

Region or Division recognition is one of the important tasks of classifying people from a specific geographical area. For our contribution, we consider eight divisions of Bangladesh named Barisal, Chittagong, Dhaka, Khulna, Mymensingh, Rajshahi, Rangpur, and Sylhet as eight labels for speakers Division recognition. The division recognition approach has a lot of applications including the detection of the speaker's geographical region, verifying the area of unknown crime suspects, and speaker recognition. Nowadays a lot of harassment and crime has been performed using the Voice over Internet Protocol (VoIP).  From the perspective of Bangladesh, a huge number of fraud calls and voice threads are performed every year.\\

In this research, we proposed a deep learning model using the Artificial Neural Network to recognize Bangladeshi speaker’s division from continuous Bengali Speech. Which will help to recognize the unknown speaker’s division in terms of crime suspect and voice fraud. It will also contribute to the Bangla Natural Language Processing and Bangla Automatic Speech Recognition community.\\

\section{Literature Review}
Recent Automatic Speech Recognition (ASR) technology is improving rapidly only for the huge number of research and development. Researchers are working on ASR technology and their implementation in real life as a huge amount of speech data is produced every day but it is not possible to store and use all this data. Biomechanics technology and applied linguistics are the most commonly used in early ASR technology. However, the modern ASR is improving because of the variety of attributes that vary speaker by speaker. Different factors that affect human speaker recognition application are described by Amino et al \cite{b1}. In terms of detecting unknown speakers a self-learning speech-controlled system for speaker identification and adaptation using an Unsupervised Speech Controlled System is performed by Herbig et al \cite{b2} -\cite{b3}. There are also a lot of speech-based speakers’ gender, height, weight, and age detection approaches done by different researchers \cite{b4} -\cite{b7}.\\

Region or Division detection is one of the important tasks of classifying people from a specific geographical area. We found some work related to speaker’s region detection using their voice. The most similar work is done by M. F. Hossain et al \cite{b8} where they tried to recognize speakers from different regions of the United Kingdom using voice based on Accent classification. They used a traditional Machine Learning algorithm on speech features extracted by MFCC and showed a maximum accuracy of 99\% with k-NN on Crowdsourced high-quality UK and Ireland English Dialect speech datasets \cite{b9}. Different accent detection is done by Joseph et al \cite{b10} for four Indian languages Bengali, Gujarati, Malayalam, and Marathi. They performed the region detection approach from different languages spoken in India based on their accent using the Dynamic Time Warping (DTW) algorithm which shows an impressive result. For the Tagalog language in the Philippines, Danao et al \cite{b11} perform a regional accent detection approach with an accuracy of 93.33\% using a Multi-Layer Perceptron (MLP) classifier.\\

Deep learning is widely used in the field of ASR as the speech is the continuous waveform of an audio signal. So sequential models like LSTM, RNN, and ANN work better for speech signals. An overview of using Artificial Neural Network (ANN) approaches used in speech recognition is reviewed by Hennebert et al \cite{b12} where some basic principles of neural networks are briefly described as well as their current applications and performances in speech recognition. In most cases, MFCC performs better than any other feature extraction algorithm in the field of ASR research. \\

Most of the voice-based region recognition applications work on Accent. We also found some accent-based voice recognition applications. We found that Mannepalli et al \cite{b13} introduced a method to identify three different accents namely Coastal Andhra, Rayalaseema, and Telangana of the Telugu language using the Nearest Neighbor Classifier and achieved 72\% accuracy. For Chinese accent detection, Long et al \cite{b14} perform a method based on RASTA – PLP algorithm extracting features known as the short-time spectrum of each speech segment and records an accuracy of 80.8\% using Naïve Bayes classifier. Zheng et al \cite{b15} propose a new combination of accent discriminative acoustic features, accent detection, and an acoustic adaptation approach for accented Chinese speech recognition. We don’t find any related papers for the Bangla Language.\\

\section{Methodology}
We tried to recognize the division of speakers from their voices. We focused on continuous speech spoken in Bangla. The whole procedure of our work has been shown in Fig 1 as a workflow.
\begin{figure}[ht]
    \centering
    \includegraphics[width=9cm]{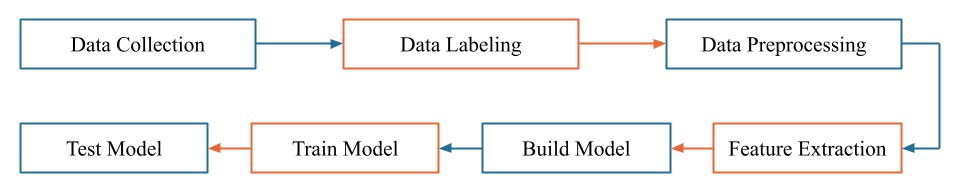}
    \caption{Overall workflow.}
\end{figure}
\subsection{Dataset}
We spent most of our time in data collection. It took more than 9 months to collect this data. We collected most of the data onsite. Some of the data were collected online through Google Forms. We have collected more than 45 hours of audio data with more than 16 thousand audio samples. Around 633 speakers (both male and female) contributed to create this dataset. We used more than 350 different scripts. Each script contains a poem, a story, and news. All the audio files were recorded in WAV format and same configuration.\\

Among 633 speakers, there were 416 male speakers and 217 female speakers. A comparison chart of male-female speakers has been provided in Fig 2.
\begin{figure}[ht]
\centerline{\includegraphics[height=6cm]{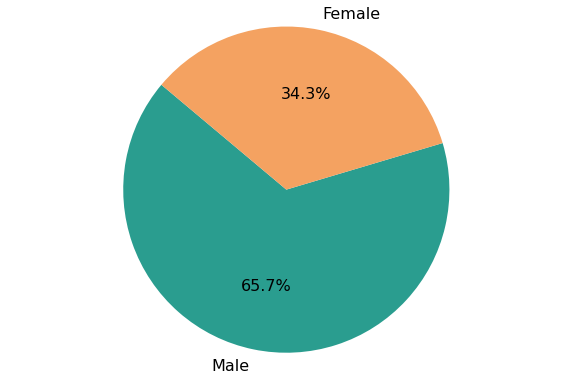}}
\caption{A ratio of male-female speakers.}
\end{figure}
There are eight divisions of Bangladesh; Barisal, Chittagong, Dhaka, Khulna, Mymensingh, Rajshahi, Rangpur, and Sylhet. We have collected data from all the divisions. While labeling the data, we considered speakers’ hometowns as their divisions. We tried to collect accent-sensitive data by suggesting speakers speak in their home division accent. We divided the dataset into eight groups and each group corresponds to one of the eight divisions. A comparison chart of total data in each division or label has been provided in Fig 3.
\begin{figure}[ht]
\centerline{\includegraphics[height=6cm]{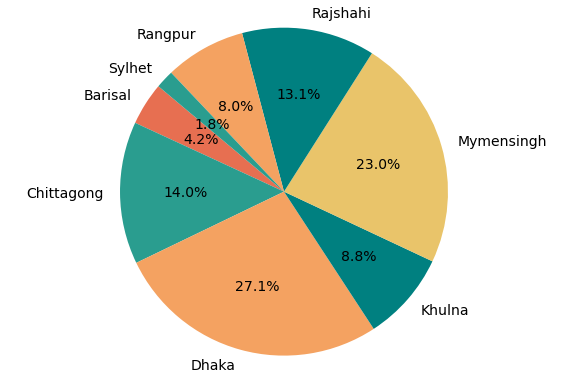}}
\caption{Amount of data at each label.}
\end{figure}

\subsection{Data Preprocessing}
After collecting data, we have done some preprocessing. We know that all the data samples needed to be measured on the same scale. So, we segmented all the audio samples in the same duration. Each audio sample was segmented into multiple chunks of 8 to 10 sec duration. Before feature extraction, we also reduced the noise from each data sample. 

\subsection{Feature Extraction}

\subsubsection{Mel Frequency Cepstral Coefficients (MFCCs)}
MFCC has always been one of the most popular features. After getting introduced in the late 70s \cite{b16}, it has overperformed other speech features for speech recognition. A huge number of experiments and developments have been made and as a result, different implementations have been proposed. Todor Ganchev, with his team, have compared the most popular four implementations for speaker recognition including the first implementation known as MFCC FB-20 implementation [17]. Among the most popular implementations, MFCC FB-40 implementation has been used in this work. The implementation steps have been discussed below.\\

\begin{itemize}
\item \textbf{Framing:} One of the best ways to get perfect features is to measure the data on the same scale. But the audio signal is not static, it's a quasi-stationary signal. This means, that it behaves like static in a short period but changes over a long period. That’s why we need to split the whole audio into some short time frame so that they behave as static and we can measure them on the same scale. The default time limit for this segmentation is 25 ms. As we said most of our data samples have a duration of 10 sec, and the sample rate is 16 kHz, after the framing, there are 400 segments or frames of each audio and each frame has (0.025*16000) = 400 samples.\\
\item \textbf{Calculating Power Spectrum:} Power spectrum gives an estimation of frequencies present in each frame. To calculate that, the Discrete Fourier Transform has been applied to each frame. There is an equation \cite{b18} for this step.

\begin{equation}
P_i(k)=\frac{1}{N}\left\lvert\sum^N_{n=1}s_i(n)h(n)e^{-j(2\div N)}\right\rvert^2\:\:\:\:1\leq k \leq K
\end{equation}

Here, P\textsubscript{i}(k) is the power spectrum of the ith frame, S\textsubscript{i}(n) is the signal representation of each sample of the ith frame, h(n) is a hamming window and K is the length of the Discrete Fourier Transform.\\
\item \textbf{Applying Mel Filtebank:} It is the most important part of MFCC implementation. In this implementation, a set of 40 equal-area filters have been used. The mathematical representation \cite{b17} of each filter is given below.
\begin{equation}
\begin{cases}
\hspace{1.45cm}0 & for\:\:k<f_{b_{i-1}}\\
\frac{2(k-f_{b_{i-1}})}{(f_{b_i}-f_{b_{i-1}})(f_{b_{i+1}}-f_{b_{i-1}})} & for\:\:f_{b_{i-1}} \leq k \leq f_{b_i}\\
\frac{2(f_{b_{i+1}}-k)}{(f_{b_{i+1}}-f_{b_i})(f_{b_{i+1}}-f_{b_{i-1}})} & for\:\:f_{b_i} \leq k \leq f_{b_{i+1}}\\
\hspace{1.45cm}0 & for\:\:k>f_{b_{i+1}}
\end{cases}
\end{equation}
Here  represents the boundary of the ith filter and k represents the k-th coefficient of the N - point DFT.\\
These 40 filters are a set of 40 vectors with some values in different positions of different vectors. The rest of the values of the vectors are zero. When each of the vectors is multiplied by the power spectrum, the sum of each vector represents the energy of the corresponding segment of the power spectrum. After this, there will be 40 energy representations for 40 filters.\\
\item \textbf{Taking the log of energies:} Now the logarithm of the energy estimation is taken for each filterbank section.\\
\item \textbf{Taking the DCT of log energies:} In the final step, a Discrete Cosine Transform (DCT) is applied to all the log energies. This represents the 40 features of the audio as energy estimation. \\
\end{itemize}

After calculating the 40 energy features, we have only taken the first 13 features. Because these first 13 features are linear features and mostly represent the signal. Figure 4 shows a comparison of MFCC feature values between each label.\\

\begin{figure}[ht]
\centerline{\includegraphics[height=6cm]{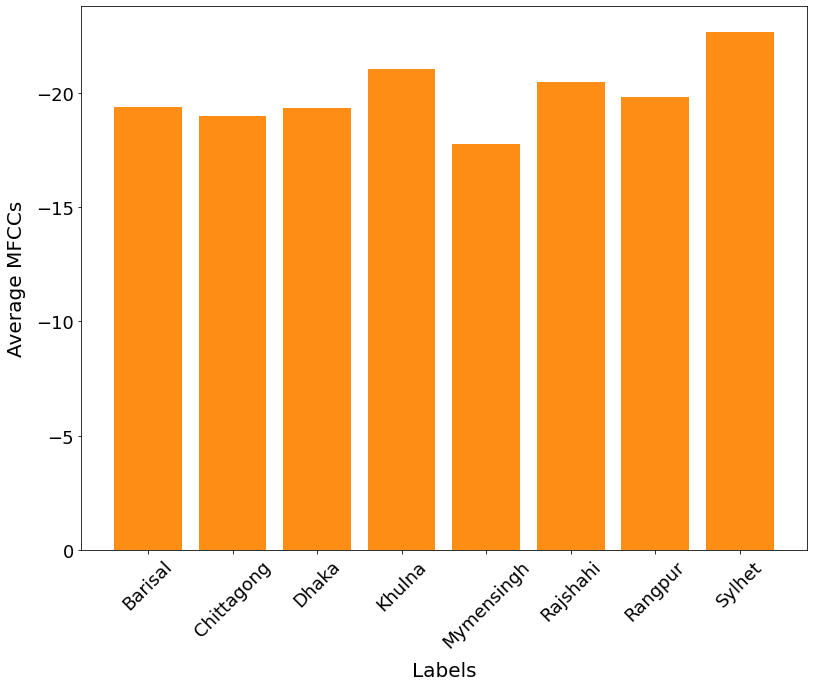}}
\caption{Average MFCCs for each label.}
\end{figure}
\subsubsection{Delta Features}
One of the major drawbacks of MFCC is it only contains the static information. But when we talk about continuous speech, it has both static and dynamic characteristics depending on the time frame. To retrieve that information, we need to use dynamic features. It has been proven that dynamic features also contain some useful information and a combination of static features and dynamic features can reduce the error rate to half \cite{b19} making the model better. So, we used dynamic features known as delta features which are calculated from MFCC by the first-order derivation. This first-order derivation gives us the same number of features we have as MFCC. We have used a combination of MFCC and delta features.

\subsection{Proposed Model}
We have experimented with different models. However, it is found that a shallow Artificial Neural Network performs better than others. The model has been described below.\\

\begin{table}[ht]
\begin{center}
\begin{tabular}{r c l}
\hline
\multicolumn{3}{l}{\textbf{Division Recognition Model}}\\
\hline
1 & : & ADAM (learning rate)\\
2 & : & For 106 iterations in all batches do: \\
3 & : & Dense (Node, Activation) \\
4 & : & Dense (Node, Activation) \\
5 & : & Dropout (Rate) \\
6 & : & Dense (Node, Activation) \\
7 & : & Dense (Node, Activation) \\
8 & : & Dense (Node, Activation) \\
9 & : & Dense (Node, Activation) \\
10 & : & end for \\
\hline
\end{tabular}
\end{center}
\end{table}

This model is a neural network of fully connected dense layers for classifying the Bengali Division. In this model, we have also used some regularization methods like batch normalization \cite{b20} and dropout \cite{b21}.

This model contains five dense layers and one output layer. Each dense layer has different numbers of hidden units or neurons. The first dense layer has 128 hidden units, the second and third dense layers have 256 hidden units, the fourth dense layer has 64 hidden units and the fifth dense layer has 32 hidden units. The ReLU (3) activation function for each dense layer except the output layer. We have used 20\% dropout twice as regularization, after the third layer and the fourth layer.
\begin{equation}
ReLU(y) = max (0, y)
\end{equation}
The output layer contains eight hidden units, the same amount as the number of labels. We have used the SoftMax (4) activation function in this layer.
\begin{equation}
\sigma(z)_{j}=\frac{e^z j}{\sum^K_{k=1} e^Z k}\;for\;j\;=\;1, ...k
\end{equation}
Figure 5 shows the proposed Division Recognition architecture.\\


\begin{figure*}[ht]
\centerline{\includegraphics[height=6cm]{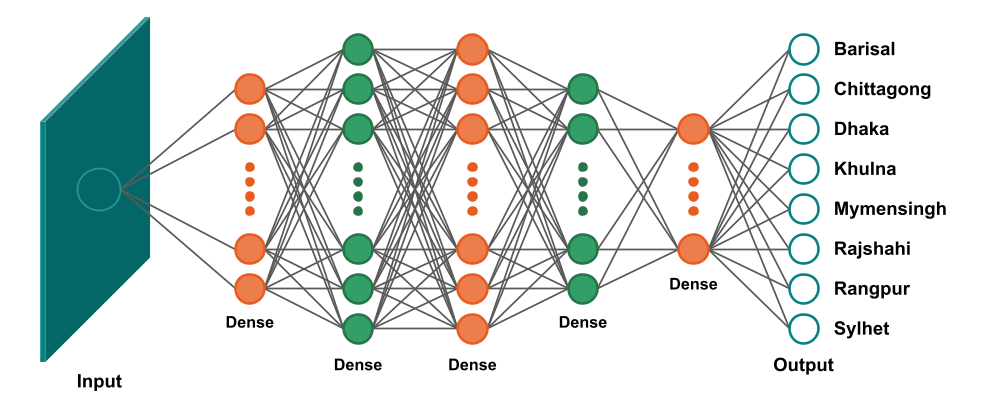}}
\caption{Architecture of Division Recognition Model.}
\end{figure*}

The model architecture summary will help us to determine the model more clearly. Table 1 shows the summary for a clear visualization of the whole model.

\begin{table}[ht]
\caption{Age Recognition Model Architecture Summary}
\begin{center}
\begin{tabular}{l c c c}
\hline
\textbf{Layer No(Type)} & \textbf{Output Shape} & \textbf{Parameter} & \textbf{Connected To} \\
\hline
(Input Layer) & 128 & 3456 - \\
1(Dense) & 256 & 33024 & Input Layer \\
2(Dense) & 256 & 65792 & 1 \\
3(Dropout) & 256 & 0 & 2 \\
4(Dense) & 64 & 16448 & 3 \\
5(Dropout) & 64 & 0 & 4 \\
6(Dense) & 32 & 2080 & 5 \\
7(Dense) & 8 & 264 & 6 \\
\hline
\multicolumn{4}{l}{Total params: 121,064}\\
\multicolumn{4}{l}{Trainable Params:121,064}\\
\multicolumn{4}{l}{Non-trainable params: 0}\\
\end{tabular}
\end{center}
\end{table}
\subsection{Optimizer and Learning Rate}\label{AA}
Researchers will minimize neural network algorithm error with the help of the Optimizer algorithm Proposed Bengali Division Classification model using Adam Optimizer \cite{b22}. Most of the researchers use it to get better performance. This Optimizer is a reform of the stochastic gradient descent algorithm. The proposed Bengali Division Classification used Adam optimizer (5) with a learning rate of 0.001.
\begin{equation}
\theta_{t+1}=\theta_t - \frac{\eta}{^1\sqrt{\hat{v}_t} + \epsilon}\hat{m}_t
\end{equation}
We calculated the error of our model using a function named categorical cross entropy (6) which is needed to optimize the algorithm. A study says that cross-entropy has outperformed other loss-calculating functions such as classification error and mean squared error etc \cite{b23}.
\begin{equation}
L_i=-\sum_j t_{i, j}\log(p_{i, j})
\end{equation}
\subsection{Training the Model}
The Bengali Division Classification model was trained on our dataset with a batch size of 128. After 35 epochs the model got almost 85\% accuracy. The optimizer converged faster by reducing the learning rate with the help of the automatic learning rate method.\\

\section{Model Performance}
The model was trained and validated on our dataset and gave a promising result on a train set, test set, and validation set. We have a total data of 16730 segmented audio as we used our 80\% data for training the model, we used 10\% for testing, and 10\% for validation.\\

The following Figure 3 shows the accuracy and loss of the division recognition model for the training and validation set.\\
\begin{figure}[ht]
\centerline{\includegraphics[width=8cm]{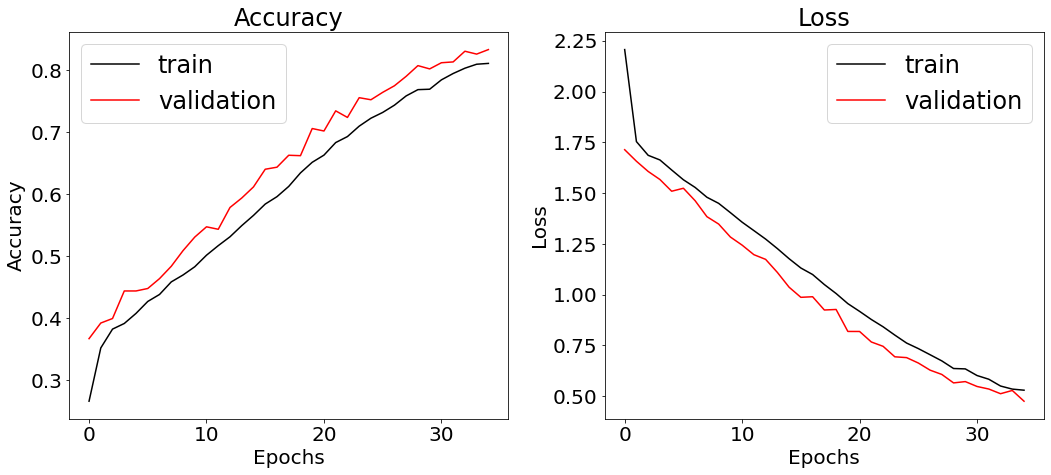}}
\caption{Accuracy and loss during training and validation of division recognition model.}
\end{figure}
After 35 epochs, the proposed model got 83.99\% accuracy on the training set and 85.44\% accuracy on a validation set of our datasets. Figure 4 shows the confusion matrix of this model so that the model performance will be clearer to all.\\
\begin{figure}[ht]
\centerline{\includegraphics[height=6cm]{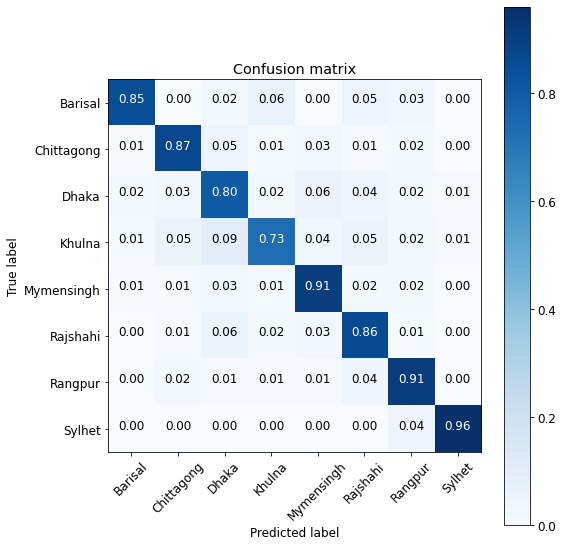}}

\caption{Confusion matrix of division recognition model.}
\end{figure}
\section{Conclusion}
Bangla is the most precious and beloved language for the Bengali people. In this research, we tried to recognize the speaker's division from Continuous Bengali Speech. We consider Barisal, Chittagong, Dhaka, Khulna, Mymensingh, Rajshahi, Rangpur, and Sylhet, the eight divisions of Bangladesh as the Geographical area. We used our dataset of 45 hours of audio containing Story, News, and Poem scripts from 633 speakers. We performed some preprocessing over our raw audio like labeling, noise reduction, and audio segmentation. We perform feature selection and use MFCC for feature extraction. We applied an Artificial Neural Network to this extracted feature and got the highest accuracy of 85\% for speaker division recognition.

\end{document}